\documentclass[twocolumn, pra]{revtex4-1}
\usepackage{amsmath}
\usepackage{graphicx}
\usepackage{amssymb}
\usepackage{latexsym}
\usepackage{verbatim,times,bbm}
\usepackage{latexsym}
\usepackage{hyperref}
\hypersetup{
     colorlinks   = true,
     citecolor    = blue,
     urlcolor = blue}

\begin{document}

\title{Comment: Inductive entanglement classification of four qubits under stochastic local operation and classical communication}

\author{Masoud Gharahi Ghahi}
\email[E-mail: ]{masoud.gharahi@gmail.com}
\affiliation{\small School of Advanced Studies, University of Camerino, 62032 Camerino, Italy}
\author{Stefano Mancini}
\email[E-mail: ]{stefano.mancini@unicam.it}
\affiliation{\small School of Science and Technology, University of Camerino, 62032 Camerino, Italy}
\affiliation{\small INFN Sezione di Perugia, I-06123 Perugia, Italy}
\date{\today}

\date{\today}

\begin{abstract}
L. Lamata et al. have presented an entanglement classification of four qubits under stochastic local operation and classical communication (SLOCC) based on inductive approach [Phys. Rev. A \textbf{75}, 022318 (2007)]. While it works well for three qubits, we show that some of the families for four qubits have overlaps and hence this approach is not valid for entanglement classification of four or more qubits under SLOCC.
\end{abstract}

\maketitle

As equivalent states perform the same tasks in quantum information processing, it is necessary to introduce an equivalent relation such that two quantum states belong to the same equivalence class. Because of non-locality of quantum entanglement, a proper equivalent relation should be local operation. Bennett et al. \cite{BPRST00} have shown that if any two states are equivalent under local operation and classical communication (LOCC), they are related by a local unitary (LU) transformation. It means that if $|\psi\rangle = U_{1} \otimes U_{2} \otimes \cdots \otimes U_{n} |\phi\rangle$, where $|\psi\rangle$ and $|\phi\rangle$ are two $n$-partite quantum states belonging to the same Hilbert space and each $U_{i}$ is a local unitary operator, one state can be obtained with certainty from the other one by means of LOCC. Thus, LOCC equivalence classes are simply the orbits under the action of LU group on the set of multipartite states. In spite of the fact that LOCC lead to a fine classification, it leads already for even two qubits systems to the existence of infinite number of LU equivalent classes. In Ref. \cite{DVC00} D\"{u}r et al. have introduced another equivalence relation based on local invertible (LI) operations. They have showed that if $|\psi\rangle = GL_{1} \otimes GL_{2} \otimes \cdots \otimes GL_{n} |\phi\rangle$, where $|\psi\rangle$ and $|\phi\rangle$ are again two $n$-partite quantum states belonging to the same Hilbert space and each $GL_{i}$ is a local invertible operator, one state can be obtained with nonzero probability from the other one by means of stochastic local operation and classical communication (SLOCC). Similarly, it turns out that SLOCC equivalence classes are the orbits under the action of LI group on the set of multipartite states. For two qubits systems there are two SLOCC classes namely, separable and entangled states and for three qubits systems there are six SLOCC classes namely, fully separable states, three biseparable states, and two inequivalent fully inseparable states GHZ and W. However, for four or more qubits there are infinite number of SLOCC classes. Hence, it is desirable to bunch the infinite number of SLOCC classes in a finite number of family with the common physical and/or mathematical properties \citep{SEDSLS16}.

Lamata et al. \citep{LLSS06} have introduced an inductive method to partition the infinite number of SLOCC classes into a finite number of families. Based on this method, they have bunched four qubits SLOCC classes into eight entanglement families (up to qubit permutation) \citep{LLSS07}. Some years later Backens has shown that the inductive approach yields ten entanglement families for four-qubit entangled states \citep{Backens17}. The main idea of this approach is to investigate the possible entanglement families of the right singular subspace of the coefficient matrix of each pure state $|\psi\rangle$ expressed in the canonical basis. In this order, one can write an $n$-qubit pure state as $|\psi\rangle=|e_{0}\rangle|\varphi_{0}\rangle+|e_{1}\rangle|\varphi_{1}\rangle$, where $|e_{0}\rangle$ and $|e_{1}\rangle$ are two linearly independent states of the $i$-th ($i=1,2,\cdots,n$) qubit, and $|\varphi_{0}\rangle$ and $|\varphi_{1}\rangle$ are the states of the rest $n-1$ qubits. In general, normalization is not needed as SLOCC operations can change the norm of the states. The entanglement families are determined by considering all combination of entanglement types of the rest $n-1$ qubits in different spanning sets for span$\{|\varphi_{0}\rangle,|\varphi_{1}\rangle\}$. Lamata et al. label the entanglement families according to the types of entangled vectors in the spanning set where $|\varphi_{0}\rangle$ and $|\varphi_{1}\rangle$ can take the values $000$, $0_{k}\Psi$, GHZ, or W. Here, $000$ denotes a fully separable state, while $0_{k}\Psi$ denotes a biseparable state where $0_{k}$ is the $k$-th single qubit in a product with an entangled state of the remaining qubits.

Here, we present two examples of four qubits states that invalidates this approach when we consider different partitioning of subsystems.

In Ref. \cite{LLSS06, LLSS07, Backens17}, the authors have considered the partition of the first qubit from the rest ($1|23$ and $1|234$) but there should be no loss of generality in choosing other partitions as they also have mentioned it in Ref. \citep{LLSS06}. For the first example, let us consider two states of the family span$\{000,0_{k}\Psi\}$ for four qubits where the partition is $1|234$
\begin{eqnarray*}
|0000\rangle+|1100\rangle+|1111\rangle &=& |0\rangle|000\rangle+|1\rangle|1\rangle(|00\rangle+|11\rangle)\, , \cr
|0000\rangle+|1101\rangle+|1110\rangle &=& |0\rangle|000\rangle+|1\rangle|1\rangle(|01\rangle+|10\rangle)\, .
\end{eqnarray*}
Now, let to consider the partition $123|4$ for the above states. So we have
\begin{eqnarray*}
|0000\rangle+|1100\rangle+|1111\rangle &=& (|00\rangle+|11\rangle)|0\rangle|0\rangle+|111\rangle|1\rangle\, , \cr
|0000\rangle+|1101\rangle+|1110\rangle &=& (|000\rangle+|111\rangle)|0\rangle+|110\rangle|1\rangle\, .
\end{eqnarray*}
As we see by changing the partition (changing the algorithm) the first state remains in the family span$\{000,0_{k}\Psi\}$ but the second one goes to the family span$\{000,\text{GHZ}\}$. We can also see this problem with two other states of the family span$\{0_{k}\Psi,0_{k}\Psi\}$ with considering the partition as $1|234$ as follow
\begin{eqnarray*}
&& |0000\rangle+|1100\rangle+\lambda_{1}|0011\rangle+\lambda_{2}|1111\rangle = \cr
&& |0\rangle|0\rangle(|00\rangle+\lambda_{1}|11\rangle)+|1\rangle|1\rangle(|00\rangle+\lambda_{2}|11\rangle)\, ,
\end{eqnarray*}
and
\begin{eqnarray*}
&& |0000\rangle+|1100\rangle+\lambda_{1}|0001\rangle+\lambda_{1}|0010\rangle \cr
&& +\lambda_{2}|1101\rangle+\lambda_{2}|1110\rangle = \cr
&& |0\rangle|0\rangle(|00\rangle+\lambda_{1}|01\rangle+\lambda_{1}|10\rangle) \cr
&& +|1\rangle|1\rangle(|00\rangle+\lambda_{2}|01\rangle+\lambda_{2}|10\rangle)\, .
\end{eqnarray*}%
But when we consider the partition as $123|4$ we have
\begin{eqnarray*}
&& |0000\rangle+|1100\rangle+\lambda_{1}|0011\rangle+\lambda_{2}|1111\rangle = \cr
&& (|00\rangle+|11\rangle)|0\rangle|0\rangle+(\lambda_{1}|00\rangle+\lambda_{2}|11\rangle)|1\rangle|1\rangle\, ,
\end{eqnarray*}
and
\begin{eqnarray*}
&& |0000\rangle+|1100\rangle+\lambda_{1}|0001\rangle+\lambda_{1}|0010\rangle \cr
&& +\lambda_{2}|1101\rangle+\lambda_{2}|1110\rangle = \cr
&& (|000\rangle+|110\rangle+\lambda_{1}|001\rangle+\lambda_{2}|111\rangle)|0\rangle \cr
&& +(\lambda_{1}|00\rangle+\lambda_{2}|11\rangle)|0\rangle|1\rangle\, ,
\end{eqnarray*}%
where the first state remains in the family span$\{0_{k}\Psi,0_{k}\Psi\}$ and the second one goes to the family span$\{0_{k}\Psi,\text{GHZ}\}$. It is worth noting that since genuine entanglement families for three qubits are merely W and GHZ with the symmetric canonical forms, there is no such a problem. Clearly this is due to the fact that these two inseparable classes are invariant under exchanging the parties and hence under changing the partition. The situation is different in the systems of four or more qubits. Indeed, from the above examples one can easily find out that there are some genuine entanglement families for four qubits for which the canonical forms are  not symmetric and these are where violation to the classification of Ref. \cite{LLSS07} takes place. Evidently, symmetric genuine entanglement families for four or more qubits are invariant under changing the partition, e.g. the family span$\{000,000\}$ and the family span$\{000,\text{W}\}$ in four qubits systems. Generally, entanglement families W and GHZ are symmetric in $n\geq3$ qubits systems.

In summary, if by changing the partition all states from one family are mapped to another family there would be no problem \citep{Pexp}, but we have seen that with the approach of Ref. \cite{LLSS06,LLSS07} it may happen that only some states go to another family. Hence, there is an overlap between some families of four-qubit entangled states and thus this approach cannot be used to exactly identify to which family a given four-qubit state belongs. Furthermore, being the approach inductive, the entanglement classification turns out to be flawed also for more than four qubits systems.


\section*{Acknowledgments}
M. Gh. acknowledges useful discussions with Pedram Karimi.



\begin{thebibliography}{99}                                                                                               
\bibitem {BPRST00} {C. H. Bennett, S. Popescu, D. Rohrlich, J. A. Smolin, and A. V. Thapliyal, {\href{https://doi.org/10.1103/PhysRevA.63.012307} {Phys. Rev. A \textbf{63}, 012307 (2000)}}.}
\bibitem {DVC00} {W. D\"{u}r, G. Vidal, and J. I. Cirac, {\href{http://dx.doi.org/10.1103/PhysRevA.62.062314} {Phys. Rev. A {\textbf{62}}, 062314 (2000)}}.}
\bibitem{SEDSLS16} {M. Sanz, I. L. Egusquiza, R. Di Candia, H. Saberi, L. Lamata, and E. Solano, {\href{https://doi.org/10.1038/srep30188} {Scientific Reports \textbf{6}, 30188 (2016)}}.}
\bibitem{LLSS06} {L. Lamata, J. Le\'on, D. Salgado, and E. Solano, {\href{https://doi.org/10.1103/PhysRevA.74.052336} {Phys. Rev. A \textbf{74}, 052336 (2006)}}.}
\bibitem{LLSS07} {L. Lamata, J. Le\'on, D. Salgado, and E. Solano, {\href{https://doi.org/10.1103/PhysRevA.75.022318} {Phys. Rev. A \textbf{75}, 022318 (2007)}}.}
\bibitem{Backens17} {M. Backens, {\href{https://doi.org/10.1103/PhysRevA.95.022329} {Phys. Rev.A \textbf{95}, 022329 (2017)}}.}
\bibitem{Pexp} {It is worth to mention that for four or more qubits, the number of SLOCC classes is infinite, indeed an uncountable infinite. Therefore, Cantor's theorem ensures that there are infinite ways to allocate them into a finite number of families. It means that there are infinite number of ways to ``classify entanglement" into families. However, most of them are meaningless classifications from the point of view of physics. Hence, even if all states from one family are mapped to another family we have a meaningless entanglement classification.}
\end{thebibliography}
\end{document}